\documentclass[doublecol]{epl2}

\usepackage[utf8]{inputenc}
\usepackage{amsmath}
\usepackage{amsfonts}
\usepackage{amssymb}
\usepackage{graphicx}
\usepackage{subcaption}
\usepackage{url}
\usepackage{multirow}
\usepackage{booktabs}
\usepackage{xcolor}
\usepackage{tabularx}
\usepackage{verbatim}

\newcommand{\be}{\begin{equation}}
\newcommand{\ee}{\end{equation}}
\newcommand{\ba}{\begin{eqnarray}}
\newcommand{\ea}{\end{eqnarray}}

\DeclareUnicodeCharacter{221A}{\ensuremath{\surd}}
\renewcommand{\arraystretch}{0.95}
\begin{document}

\title{Probing Vector-Like Quarks at a Future Muon-Proton Collider}
\shorttitle{Vector-Like Quarks at a Future Muon-Proton Collider}

\author{
Ijaz Ahmed\inst{1}\thanks{Corresponding author: \texttt{ijaz.ahmed@fuuast.edu.pk}} \and
M. Tayyab Javaid\inst{1} \and
Haroon Sagheer\inst{2} \and
M. Danial Farooq\inst{1} \and
Jamil Muhammad\inst{3}\thanks{\texttt{mjamil@konkuk.ac.kr}}}

\shortauthor{Ijaz Ahmed \etal}

\institute{
\inst{1} Federal Urdu University of Arts, Science and Technology, Islamabad, Pakistan\\
\inst{2} Riphah International University, Islamabad, Pakistan\\
\inst{3} Sang-Ho College \& Department of Physics, Konkuk University, Seoul 05029, South Korea
}
\pacs{12.60.-i} {Models beyond the Standard Mode}
\pacs{14.65.Jk} {Other quarks (e.g., 4th generations, heavy partners)}
\abstract{This study examines the discovery potential of a singly produced vector-like top quark ($T$) at a future muon--proton collider with $\sqrt{s} = 5.29$, 6.48, and 9.16~TeV, using a model-independent effective Lagrangian consistent with CKM and electroweak constraints. The $T$ quark decays mainly into $Wb$; the production cross-section scales as $\kappa^2$, peaks at 9.16~TeV, and falls rapidly above 3~TeV owing to PDF and phase-space effects. The hadronic channel offers higher event rates and the leptonic channel cleaner backgrounds; at $m_T=3$~TeV and 3000~fb$^{-1}$ the hadronic significance is $21.86\sigma$ versus $3.75\sigma$ for the leptonic one, driven by the $W$ branching ratios ($\approx 67\%$ hadronic vs.\ $\approx 11\%$ per leptonic flavor). The $H_T$ discriminator in the hadronic channel rejects background better than missing-energy metrics in the leptonic final state. Boosted Decision Trees and Multi-Layer Perceptrons were applied at $\sqrt{s} = 9.16$~TeV, $m_T=3000$~GeV; the MLP outperforms the BDT at all luminosities, with a hadronic purity gain of about 2.62. Overall, a muon--proton collider can probe $T$ quark masses up to approximately 3.5~TeV, surpassing current limits.} 






\maketitle

\section{Introduction}
Many Standard Model (SM) extensions, including Little Higgs \cite{Randall1999, ArkaniHamed2002}, composite Higgs \cite{Agashe2005, Kaplan1984, Contino2007} and two Higgs doublet models \cite{Arhrib2018, Branco2012}, predict TeV-scale vector-like quarks (VLQs) \cite{Das2023, Cacciapaglia2023}, whose chiral components transform identically under the electroweak (EW) symmetry \cite{AguilarSaavedra2013}. VLQs may stabilize the EW vacuum \cite{Xiao2014, Joglekar2012, Benbrik2014}, address CKM unitarity \cite{Workman2022, Belfatto2021, Cheung2020} and explain anomalies \cite{Abdullah2023}, including the $W$-mass discrepancy \cite{Cao2022, Aaltonen2022, ATLAS2024, arXiv:2412.13872}. The FCC-hh is projected to probe pair-produced VLQs up to 6 TeV \cite{FCC_CDR}.

The LHC has seen no VLQ signal; pair-produced VLQs are excluded up to about 1.4--1.5 TeV \cite{ATLAS2022, CMS2025, ATLAS2025}. At higher masses QCD pair production falls rapidly and EW single production dominates \cite{AguilarSaavedra2013, Hernandez2024, Vignaroli2014}. $pp$ colliders suffer from large QCD backgrounds, while $ep$ colliders such as the LHeC are cleaner but limited in energy \cite{Bruening2020, LHeC2023}. A muon-proton ($\mu p$) collider \cite{IMCC2023} reaches higher energies than $ep$ machines owing to the low synchrotron radiation of muons, with a cleaner environment than $pp$ \cite{Caldwell2016, Acar2017, Long2023}. We study $\sqrt{s}=5.29$, 6.48 and 9.16~TeV, from 1--3 TeV muon beams on 7 TeV (LHC) and 14 TeV (FCC-p) protons, via $\sqrt{s} = 2\sqrt{E_\mu E_p}$.

The decay $T \to Wb$ often dominates over $T\to Zt, Ht$ \cite{DeSimone2013} and its high-$p_T$ $b$-quark plus $W$ signature discriminates well against the background \cite{Song2025}. We assess the discovery potential of singly produced $T$ quarks in this channel, comparing with $pp$ and $ep$ colliders. Section II defines the model, Section III the simulation; the remaining sections present the hadronic and leptonic channels, significances and discovery reach.

\section{The VLQ model and single production of vector-like T quark}
The effective Lagrangian describing single production and decay of the vector-like top partner $T$ is
\begin{equation}
\begin{aligned}
\mathcal{L}_{T}^{\text{single}} =\ & \kappa_W V^{4i}_{L/R} \frac{g}{\sqrt{2}} \left[ \overline{T}_{L/R} W_{\mu}^{+} \gamma^{\mu} d_{i\,L/R} \right] \\
&+ \kappa_Z V^{4i}_{L/R} \frac{g}{2c_W} \left[ \overline{T}_{L/R} Z_{\mu} \gamma^{\mu} u_{i\,L/R} \right] \\
&- \kappa_H V^{4i}_{L/R} \frac{M}{v} \left[ \overline{T}_{R/L} H u_{i\,L/R} \right] + \text{h.c.}
\label{Eq:lang}
\end{aligned}
\end{equation}
Here $M$ is the $T$ mass, $V^{4i}_{L/R}$ the mixing with SM generation $i$, and $\kappa_{W,Z,H}$ the $W$, $Z$, Higgs coupling strengths; this extends Ref.~[8] with Higgs interactions \cite{Das2023} and couplings to all three generations. For $\kappa_W=\kappa_Z=\kappa_H=1$ the asymptotic ($M\to\infty$) branching fractions are 50\% to $W$ and 25\% each to $Z$ and $H$ (Goldstone equivalence); related work includes $T\to tZ$ with invisible $Z$ at the 14~TeV LHC \cite{Li2023}. We define an effective coupling $g^*$ so that $\sigma\propto (g^*)^2$, with benchmarks in the perturbative regime consistent with EW precision and CKM constraints \cite{Cheung2020, Crivellin2020, Branco2021}, i.e.\ $|V_{Lb}| \lesssim 0.1$.

Single $T$ production proceeds via $\mu p \to \nu_{\mu} T b$ through $t$-channel $W$ exchange between the muon and a quark inside the proton (Fig.~1).
\begin{figure}[h!]
    \centering
    \includegraphics[width=0.5\columnwidth]{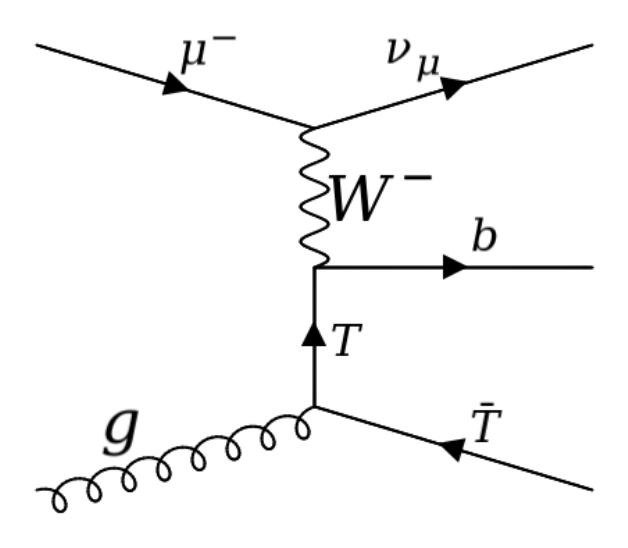}
    \caption{\textbf{Single $T$-quark production.} Leading-order (LO) Feynman diagram for the process $\mu p \to \nu_\mu T b$, including the subsequent decay channel of the $T$ quark.}
    \label{fig:production_mechanism}
\end{figure}
\begin{figure*}[t]
    \centering
   \begin{subfigure}[t]{0.27\textwidth}
\centering
\includegraphics[width=\linewidth]{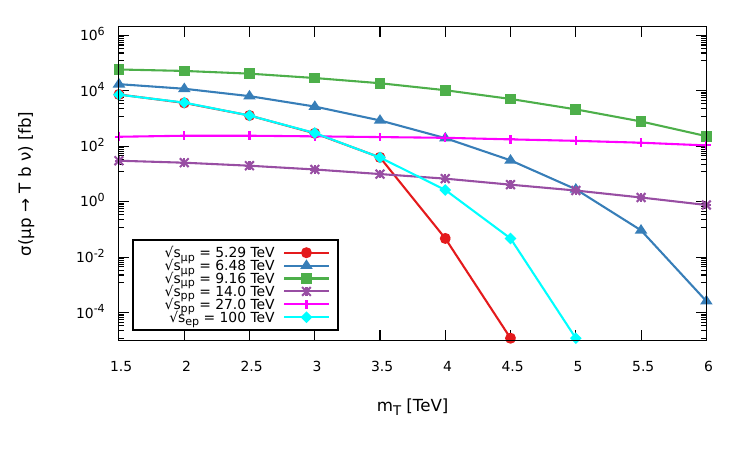}
\caption{}
\end{subfigure}\hspace{0.04\textwidth}
    \begin{subfigure}[t]{0.27\textwidth}
\centering
\includegraphics[width=\linewidth]{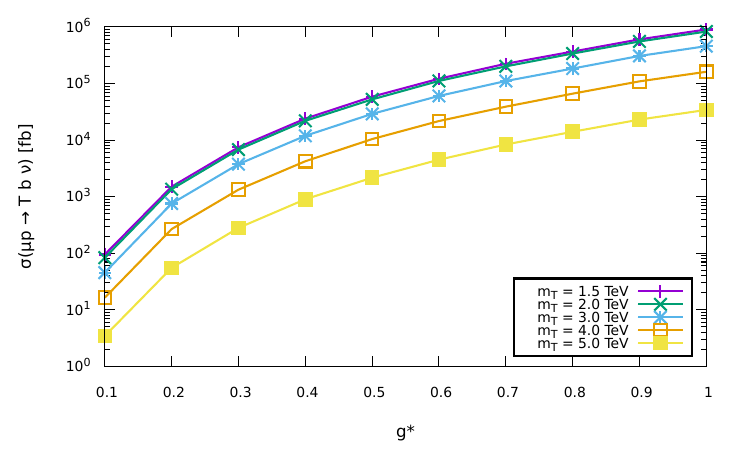}
\caption{}
\end{subfigure}
    \caption{(a) Production cross-section $\sigma(\mu p \to T b \nu)$ as a function of the VLQ mass $m_T$ for various benchmark center-of-mass energies, including comparisons with $pp$ and $ep$ collider prospects. (b) Cross-section dependence on the effective coupling $g^*$, demonstrating the characteristic quadratic scaling $\sigma \propto (g^*)^2$ essential for single production sensitivity studies.}
    \label{fig:sigma4}
\end{figure*}

Figure 2 (Left) shows the cross-section versus $m_T$ for $pp$, $ep$ and $\mu p$ colliders. The 9.16~TeV $\mu p$ machine extends the mass reach well beyond the 5.29~TeV one. The 27~TeV $pp$ collider has the largest cross-section with a flat curve, probing low Bjorken-$x$ ($x\approx0.05$--$0.22$) where PDFs are abundant \cite{Ball}, whereas the 9.16~TeV $\mu p$ collider probes $x\approx0.65$ where PDF suppression is exponential. The $\mu p$ rate exceeds even a 100~TeV $ep$ colliders since muons, about 207 times heavier than electrons, have negligible synchrotron and beamstrahlung losses ($\propto m^{-4}$) \cite{Delahaye} and deliver their full energy to the hard process; for $m_T>3.5$~TeV, however, the 5.29~TeV machine nears its kinematic limit. Figure 2 (Right) shows the $(g^*)^2$ scaling, consistent with the narrow width approximation.

\section {SIMULATION FRAMEWORK AND ANALYSIS SETUP}
Signal and backgrounds were generated at LO with a standard Monte Carlo chain: the Lagrangian of Eq.~(1) was implemented in \textbf{FeynRules} \cite{Alloul2014} (UFO output), events were generated with \textbf{MadGraph5\_aMC@NLO} \cite{Alwall2014}, showered and hadronized with \textbf{PYTHIA 8} \cite{Sjostrand2015}, and passed through \textbf{Delphes 3} \cite{deFavereau2014} with a card for lepton-hadron environments. Jets use anti-$k_t$ with $R=0.4$ in \textbf{FastJet} \cite{Cacciari2012}.

\section{Branching Ratio of VLQ T}
Figure 3 (a) shows that $T\to Wb$ dominates ($\approx 35\%$) from 600~GeV to 6~TeV, motivating our focus on $Wb$. The $T\to tZ$ and $T\to tH$ modes converge to equal rates ($\approx 16\%$) at high mass, as expected from the Goldstone Boson Equivalence Theorem \cite{Chanowitz}.

\section{Decay Width of VLQ T}
\begin{figure*}[t]
\centering
\begin{subfigure}[t]{0.27\textwidth}
\centering
\includegraphics[width=\linewidth]{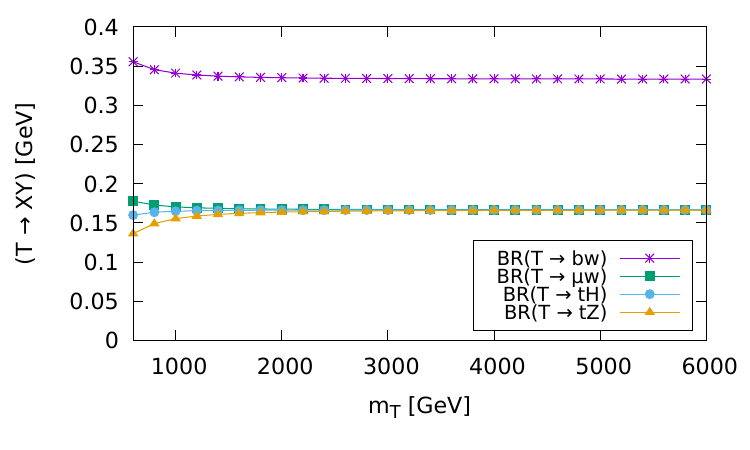}
\caption{}
\end{subfigure}\hspace{0.04\textwidth}
\begin{subfigure}[t]{0.27\textwidth}
\centering
\includegraphics[width=\linewidth]{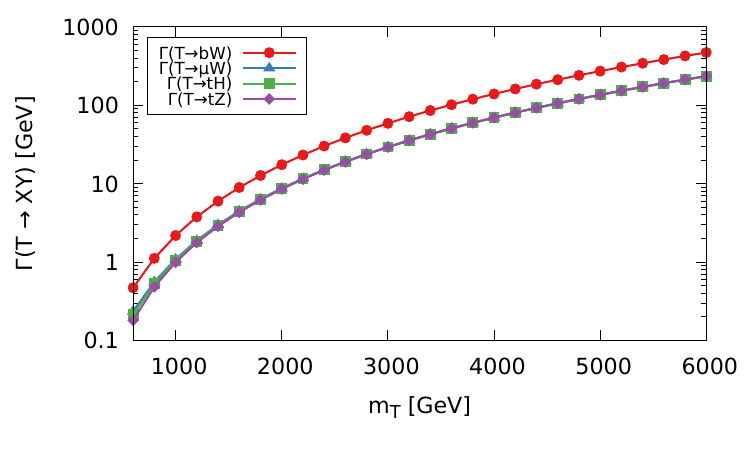}
\caption{}
\end{subfigure}
\caption{(a) Branching ratios of the vector-like $T$ quark into Standard Model final states. The dominance of the charged-current decay $T \to bW$ consistently possesses the largest partial width. Since $T$ carries a charge of $+2/3$, the decay to a $b$-quark ($-1/3$) and a $W^+$ boson ($+1$) naturally satisfies charge conservation. (b) Partial decay widths $\Gamma$ as a function of mass $m_T$. The convergence of $tZ$ and $tH$ channels at high masses illustrates the Goldstone Boson Equivalence Theorem.}
\label{fig:PDW.pdf}
\end{figure*}
Figure 3 (b) shows the partial widths $\Gamma$ for $m_T$ from 600 GeV to 6 TeV, with three features:
\begin{itemize}
\item \textit{Power-law growth:} all widths grow rapidly, $\Gamma \propto m^{3}_{T}$, as phase-space suppression vanishes and longitudinal gauge boson polarizations dominate \cite{DeSimone2013}.
\item \textit{Charged-current dominance:} $T\to Wb$ has the largest width at all masses, since $\kappa_W$ exceeds the neutral-current couplings by a factor of $\sqrt{2}$ in the singlet model.
\item \textit{Goldstone equivalence:} $T\to tH$ and $T\to tZ$ differ slightly at low mass (phase space) but converge for $m_T\gg m_Z,m_H$, as $\Gamma(T \to tZ) \approx \Gamma(T \to tH)$ \cite{Chanowitz}.
\end{itemize}
The total width is thus dominated by $Wb$, justifying our exclusive focus on this channel. The dominant SM backgrounds that mimic the signal are shown in Fig.~4 \cite{Alwall2014, Sjostrand2015}.
\begin{figure*}[t]
    \centering
{\includegraphics[width=0.24\textwidth]{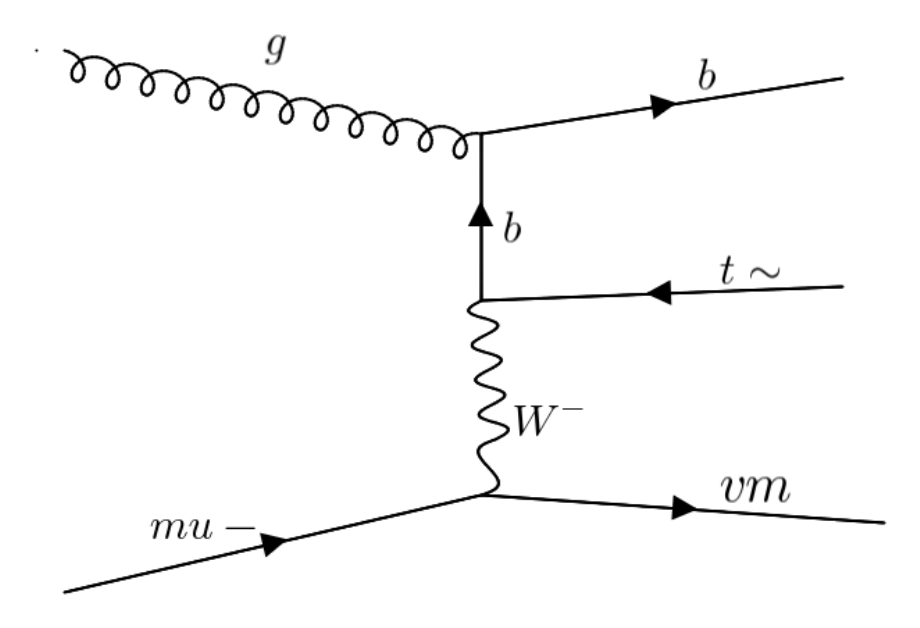}} \hspace{0.04\textwidth}
{\includegraphics[width=0.24\textwidth]{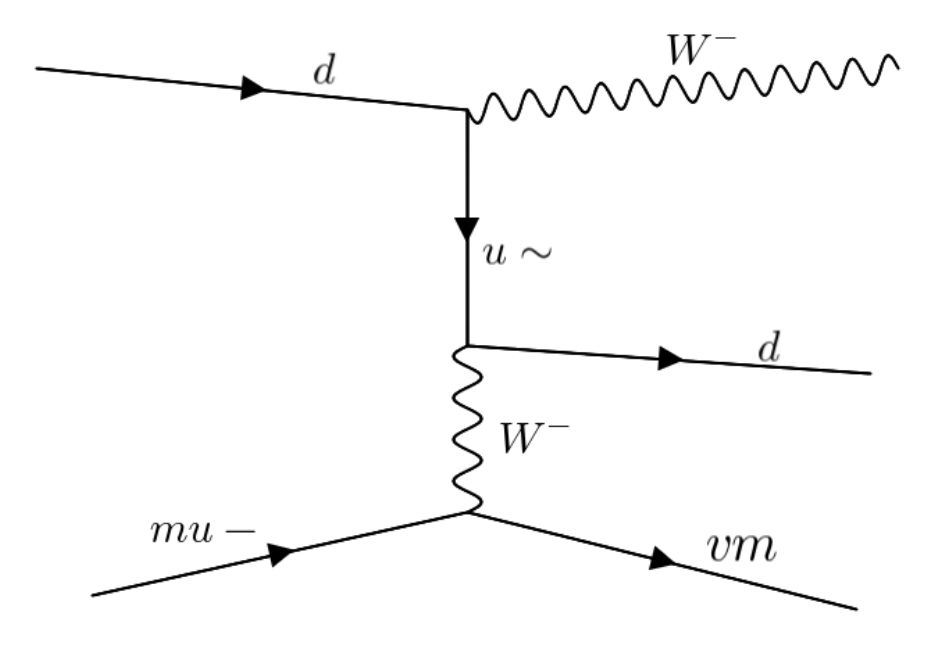}}
  \caption{Representative Feynman diagrams for the dominant Standard Model background processes leading to leptonic final states, including 
$\mu p \rightarrow \nu_\mu\,\bar{t}\, b$ and 
$\mu p \rightarrow \nu_\mu\, W^-\, d$ production, 
which mimic the signal topology.}
   \label{fig:bkglep}
\end{figure*}

\begin{table}[tbp]
\scriptsize\setlength{\tabcolsep}{2pt}\renewcommand{\arraystretch}{0.85}
\centering
\caption{Cumulative signal efficiencies and predicted event counts for various VLQ-$T$ masses ($m_T$ in GeV) at $\sqrt{s} = 9.16$ TeV. The production cross-sections $\sigma$ [fb] are shown in the first row, and the final row represents the total predicted events ($N$) for $\mathcal{L}_{\text{int}} = 3000$ fb$^{-1}$.}
\label{tab:signal_efficiency}

\resizebox{\columnwidth}{!}{
\begin{tabular}{|l|c|c|c|c|c|c|c|c|} 
\hline
\toprule
\textbf{Selection Cuts} & \textbf{$m_T=1500$} & \textbf{$m_T=2000$} & \textbf{$m_T=2500$} & \textbf{$m_T=3000$} & \textbf{$m_T=3500$} & \textbf{$m_T=4000$} & \textbf{$m_T=4500$} & \textbf{$m_T=5000$} \\
 & \textbf{GeV} & \textbf{GeV} & \textbf{GeV} & \textbf{GeV} & \textbf{GeV} & \textbf{GeV} & \textbf{GeV} & \textbf{GeV} \\  \hline
\midrule
\textbf{Cross-section $\sigma$ [fb]} & 594.9 & 575.5 & 502.8 & 425.0 & 325.5 & 230.5 & 136.9 & 70.1 \\ 
\hline
\midrule
$\eta(j) > -4.0$ & 0.9999 & 0.9999 & 1.0000 & 1.0000 & 1.0000 & 1.0000 & 0.9988 & 1.0000 \\
$\eta(j) < 7.0$ & 0.9993 & 0.9995 & 0.9998 & 0.9999 & 1.0000 & 1.0000 & 0.9988 & 1.0000 \\
$P_T(j) > 300$ GeV & 0.9083 & 0.9499 & 0.9692 & 0.9763 & 0.9842 & 0.9878 & 0.9900 & 0.9918 \\
$N(j) \geq 5$ & 0.6861 & 0.7007 & 0.6889 & 0.6538 & 0.6433 & 0.6069 & 0.5592 & 0.3977 \\
$\eta(b) > -2.0$ & 0.6845 & 0.6997 & 0.6883 & 0.6529 & 0.6428 & 0.6064 & 0.5588 & 0.3977 \\
$\eta(b) < 5.0$ & 0.6747 & 0.6888 & 0.6792 & 0.6464 & 0.6374 & 0.6037 & 0.5584 & 0.3977 \\
$P_T(b) > 20$ GeV & 0.4041 & 0.4130 & 0.4070 & 0.3916 & 0.3819 & 0.3594 & 0.3277 & 0.2207 \\
$N(b) > 2$ & 0.2124 & 0.2089 & 0.2079 & 0.1946 & 0.1839 & 0.1702 & 0.1478 & 0.0884 \\
$MET \leq 700$ GeV & 0.2122 & 0.2086 & 0.2076 & 0.1943 & 0.1832 & 0.1690 & 0.1462 & 0.0869 \\
$T_{HT} \geq 800$ GeV & 0.2084 & 0.2069 & 0.2067 & 0.1934 & 0.1824 & 0.1682 & 0.1457 & 0.0866 \\
$\Delta R(b_1, b_2) > 2.5$ & 0.1232 & 0.1208 & 0.1202 & 0.1088 & 0.1018 & 0.0883 & 0.0735 & 0.0447 \\
$\Delta R(b_1, b_2) < 4.0$ & 0.1144 & 0.1097 & 0.1085 & 0.0971 & 0.0917 & 0.0794 & 0.0659 & 0.0384 \\
\midrule
\hline
\textbf{Total Events} & \textbf{204256} & \textbf{189445} & \textbf{163674} & \textbf{123806} & \textbf{89550} & \textbf{54805} & \textbf{27065} & \textbf{8081} \\ \hline
\bottomrule
\end{tabular}
}
\end{table}

\section{Hadronic Analysis Section}
We use a cut-based analysis; Table I gives the cumulative signal efficiencies. The efficiency degrades from 1.5 to 5~TeV because the boosted decay products merge into ``fat jets'' and fail the multiplicity cut $N_{jet}\geq 5$ \cite{Abdesselam}. Requiring a $b$-tag suppresses $W+$jets and QCD multi-jet backgrounds, and the $\Delta R(b_1,b_2)$ cuts exploit the collimated topology. Large-radius jets with substructure variables would add $W$-tagging power \cite{Butterworth} and are left for future work.

Table II gives the background efficiencies for $m_T\simeq1500$~GeV: the $H_T\ge800$~GeV, $\Delta R$ and $p_T$ cuts reduce the QCD multi-jet ($vjj$) background by orders of magnitude. Table III shows that at $m_T=1.5$~TeV, $S/B\approx13$ and $S/\sqrt{B}\approx660$, but at 5~TeV $S/B$ drops to $6.3\times10^{-9}$; the decline for $m_T\ge3$~TeV follows from PDF fall-off at high Bjorken-$x$ and phase-space suppression.

\begin{table}[tbp]
\scriptsize\setlength{\tabcolsep}{2pt}\renewcommand{\arraystretch}{0.85}
\centering
\caption{Cumulative efficiency and predicted event counts for Standard Model background processes at $\sqrt{s}=9.16$ TeV. Cross-sections $\sigma$ [fb] are provided in the first row, and Total Predicted Events are scaled to $\mathcal{L}_{int} = 3000\text{ fb}^{-1}$.}
\label{table:2_revised}
\resizebox{\columnwidth}{!}{%
\begin{tabular}{|l|c|c|c|c|}
\hline
\textbf{Selection Cuts} & \textbf{$\nu tb$} & \textbf{$\nu W j$} & \textbf{$\nu Z j$} & \textbf{$\nu jj$} \\ \hline
\hline
\textbf{Cross-section $\sigma$ [fb]} & 50.0 & 50.2 & 42.5 & 51.6 \\ \hline
\hline
$\eta(j) > -4.0$ & 0.905412 & 0.945382 & 0.951468 & 0.95042 \\
$\eta(j) < 7.0$ & 0.905412 & 0.945382 & 0.951468 & 0.95042 \\
$P_T (j) > 300$ GeV & 0.00501 & 0.227075 & 0.06010 & 0.0159 \\
$N(j) \ge 5.0$ & 0.00449 & 0.080277 & 0.028536 & 0.00667 \\
$\eta(b) > -2.0$ & 0.00355 & 0.067896 & 0.024517 & 0.00560 \\
$\eta(b) < 5.0$ & 0.00355 & 0.067896 & 0.024517 & 0.00560 \\
$P_T (b) > 20$ GeV & 0.00274 & 0.056644 & 0.01983 & 0.00441 \\
$N(b) > 2.0$ & 0.00163 & 0.00566 & 0.00143 & 0.00028 \\
$MET \le 700$ GeV & 0.00161 & 0.00562 & 0.00140 & 0.00026 \\
$T_{HT} \ge 800$ GeV & 0.00087 & 0.00452 & 0.00039 & 0.00008 \\
$\Delta R(b_1, b_2) > 2.5$ & 0.00059 & 0.00247 & 0.00017 & 0.00004 \\
$\Delta R(b_1, b_2) < 4.0$ & 0.00059 & 0.00246 & 0.00016 & 0.00004 \\ \hline
Total Pred. Events & 88.5 & 370.5 & 20.4 & 6.2 \\ \hline
\end{tabular}
}
\end{table}
\begin{table}[htbp]
\scriptsize\setlength{\tabcolsep}{2pt}\renewcommand{\arraystretch}{0.85} 
\caption{All event counts represent the total predicted events for an integrated luminosity of $\mathcal{L}_{int} = 3000 \text{ fb}^{-1}$.}
\centering
\resizebox{\columnwidth}{!}{%
\begin{tabular}{|c|c|c|c|c|c|c|}
\hline
\textbf{$m_T$}  & \textbf{Signal} & \textbf{Background} & \textbf{S/B} 
& \textbf{$S/\sqrt{B}$} 
& \textbf{$S/(S+B)$} 
& \textbf{$S/\sqrt{S+B}$} \\
\hline
1500 & 43518 & 3338.5 & 13.035  & 660.0460 & 0.916  & 195.04 \\
2000 & 21064  & 3652.3 & 5.768 & 323.91495 & 0.856  & 130.700  \\
2500 & 6672  & 3569.3 & 1.8695   & 110.46450  & 0.646  & 66.660   \\
3000 & 1355.6   & 4282.5 & 0.3165  & 21.86502   & 0.243  & 18.666   \\
3500 & 177.2   & 3923.3 & 0.04518  & 2.7962  & 0.045  & 2.985   \\
4000 & 10.64   &  4177.2 & 0.002548  &  0.170898    & 0.00259  & 0.180   \\
4500 & 0.157   &  3608.3 & 4.35e-05& 2.47e-03  &  4.33e-05 & 0.00247 \\
5000 & 2.42e-05  & 3836.4  &  6.30e-09 & 3.65e-07  & 6.10e-09 & 3.76e-07 \\
\hline
\end{tabular}
}
\end{table}

Signal jets are centrally produced (peaked at $\eta=0$, with $\eta=-\ln[\tan(\theta/2)]$ approximating the rapidity $y$ for $p\gg m$), unlike forward-directed background jets.
\begin{figure}[t]
    \centering
    \includegraphics[width=0.62\columnwidth]{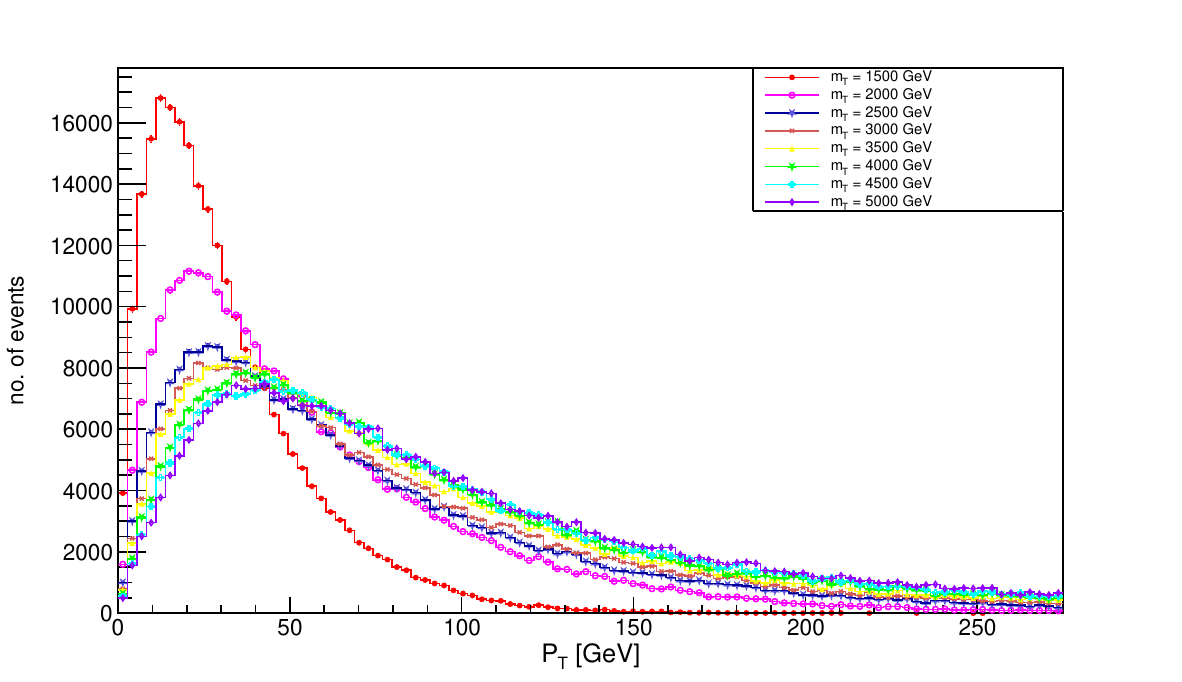}
    \caption{The transverse momentum ($p_T$) distribution of the VLQ-T for different mass benchmarks. The shift of the peak toward higher values for heavier masses illustrates the "hardening" of the signal spectrum used for background rejection.}
    \label{fig:placeholder}
\end{figure}
\begin{figure*}[t]
\centering
 \begin{subfigure}[t]{0.27\textwidth}
\centering
\includegraphics[width=\linewidth]{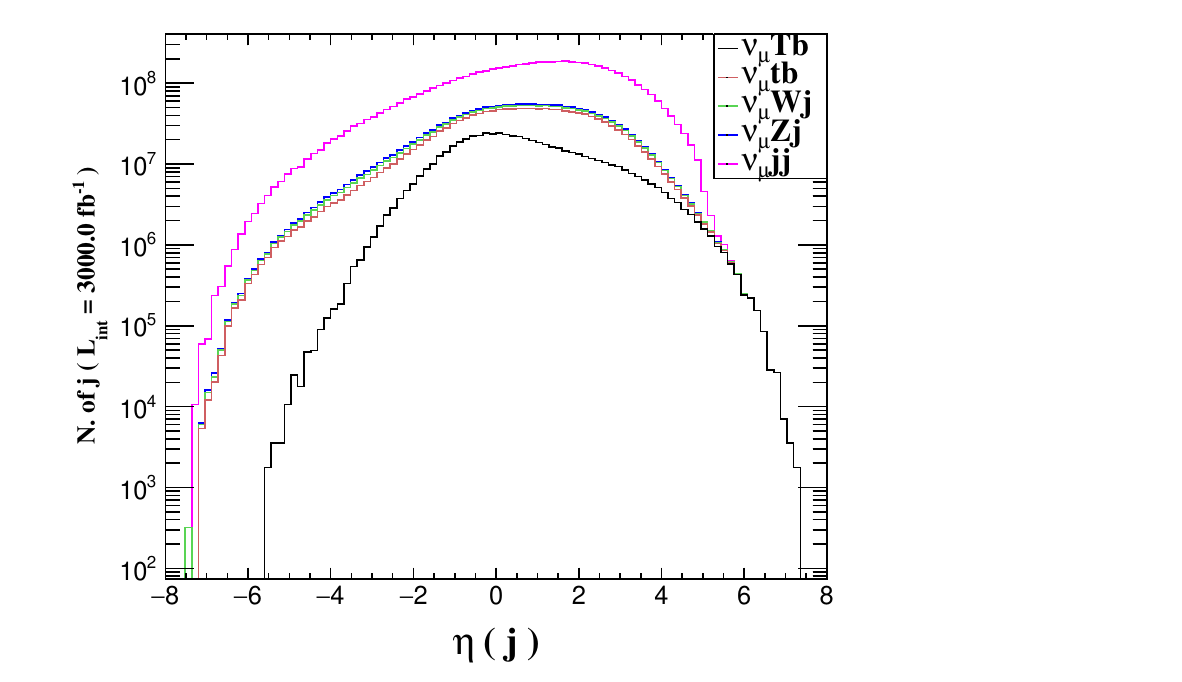}
\caption{}
\end{subfigure}\hfill
  \begin{subfigure}[t]{0.27\textwidth}
\centering
\includegraphics[width=\linewidth]{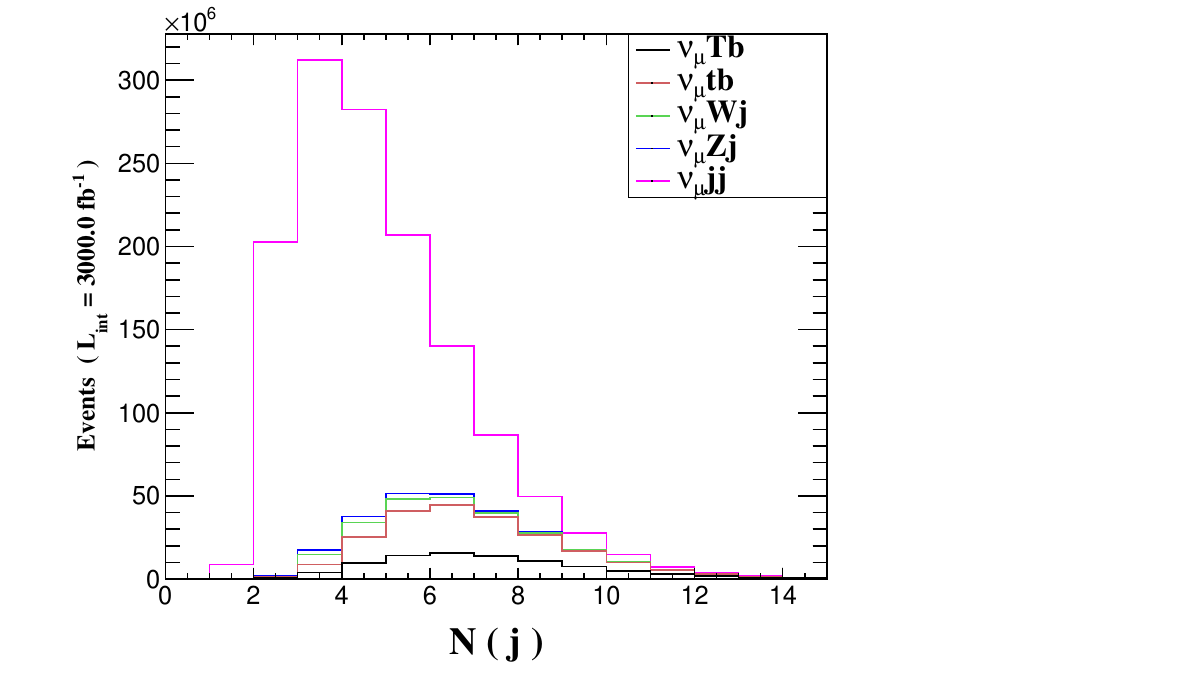}
\caption{}
\end{subfigure}\hfill
  \begin{subfigure}[t]{0.27\textwidth}
\centering
\includegraphics[width=\linewidth]{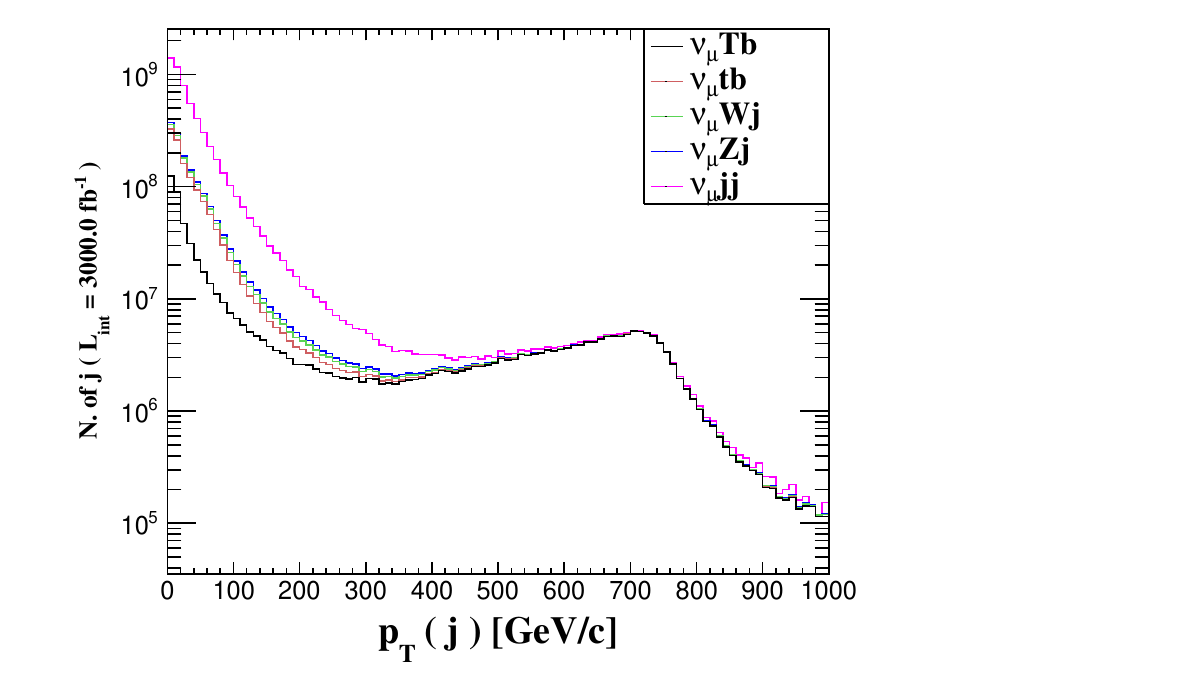}
\caption{}
\end{subfigure}
    \caption{Kinematic distributions for the hadronic channel at $\sqrt{s} = 9.16$~TeV for a benchmark mass $m_T = 1.5$~TeV and dominant SM backgrounds: 
    \textbf{(a)} Pseudorapidity ($\eta$) of all jets, showing central signal production; 
    \textbf{(b)} Jet multiplicity $N(j)$, indicating the high-occupancy jet environment; and 
    \textbf{(c)} Transverse momentum $p_T$ of reconstructed jets. 
    Distributions are shown for an integrated luminosity of 3000~fb$^{-1}$.}
    \label{fig:etajf}
\end{figure*}

\section{KINEMATIC DISTRIBUTIONS (HADRONIC)}
Figure 5 shows that the $p_T$ peak of the VLQ-T shifts upward with mass, which separates the signal from softer backgrounds. In Fig.~6, signal jets are central whereas background jets are often forward or from ISR (6a); the signal $T\to Wb\to jjb$ peaks at 5--6 jets, above $W+$jets (6b); and signal jets carry much higher $p_T$, validating the $p_T>300$~GeV cut (6c).

Figure 7(a) shows $E_T^{miss}$, whose tail arises from resolution and semi-leptonic $b$ decays, and Fig.~7(b) the total hadronic transverse energy $H_T=\sum|p_T^{jets}|$, the most powerful discriminator. The similar peak positions of signal and background result from cut-sculpting: at least 5 jets with $p_T>300$~GeV force $H_T$ to peak near 1500~GeV. The signal nevertheless keeps a harder tail, whereas backgrounds fall exponentially. We estimate a 10--15\% systematic uncertainty on the significances from the generic detector simulation. This similarity motivates multivariate classifiers (BDT, MLP).

Figure 8(a) shows the reconstructed $m_{bjj}$ with a clear peak at the generated mass, and Fig.~8(b) the significance versus VLQ mass for 500, 1000 and 3000~fb$^{-1}$, from the asymptotic formula \cite{Cowan, M.J.Basso}.
\begin{figure*}[t]
    \centering
    \begin{subfigure}[t]{0.27\textwidth}
\centering
\includegraphics[width=\linewidth]{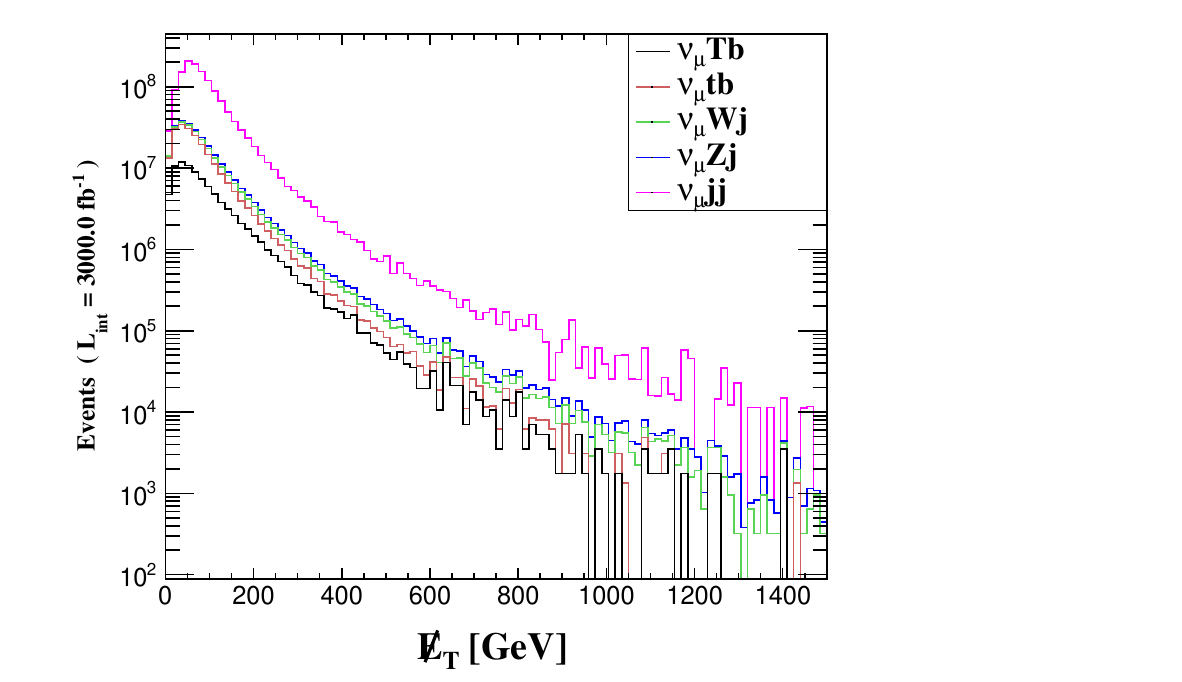}
\caption{}
\end{subfigure}\hspace{0.04\textwidth}
    \begin{subfigure}[t]{0.27\textwidth}
\centering
\includegraphics[width=\linewidth]{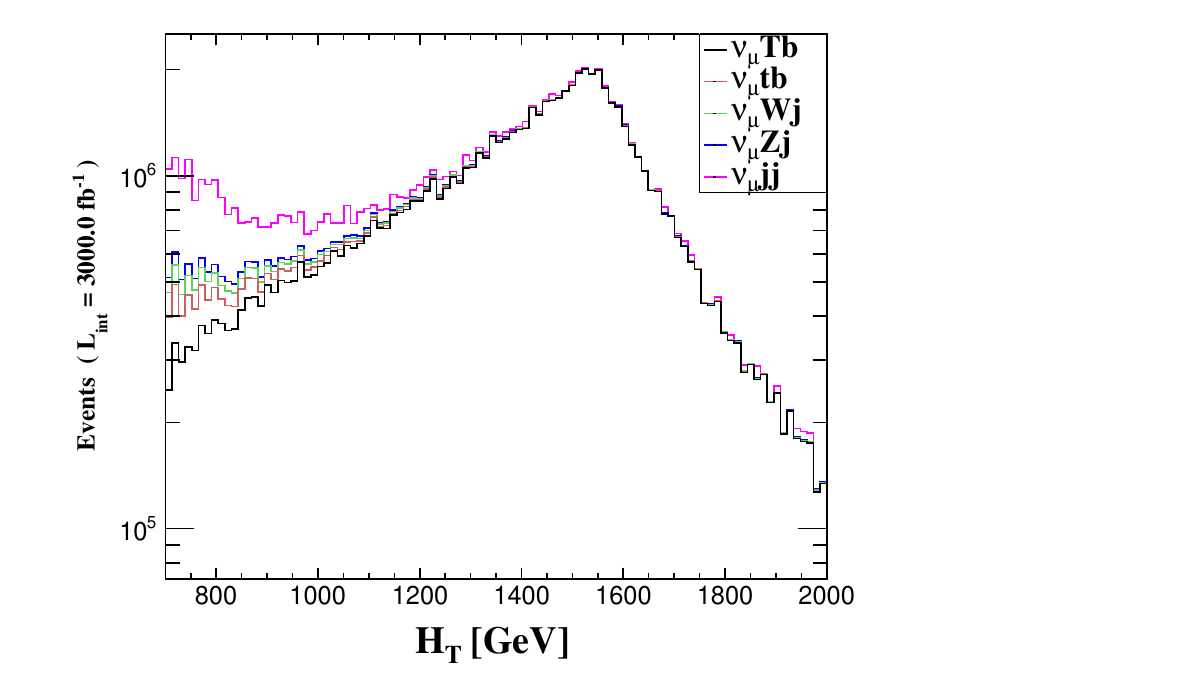}
\caption{}
\end{subfigure}
    \caption{(a) Missing transverse energy ($E_T^{miss}$) distribution for signal and background in the hadronic channel. (b) Total hadronic transverse energy ($H_T = \sum |p_T^{jets}|$), highlighting its role as a primary discriminator for multi-TeV heavy quarks.}
    \label{fig:missing-et}
\end{figure*}
\begin{figure*}[t]
    \centering
  \begin{subfigure}[t]{0.27\textwidth}
\centering
\includegraphics[width=\linewidth]{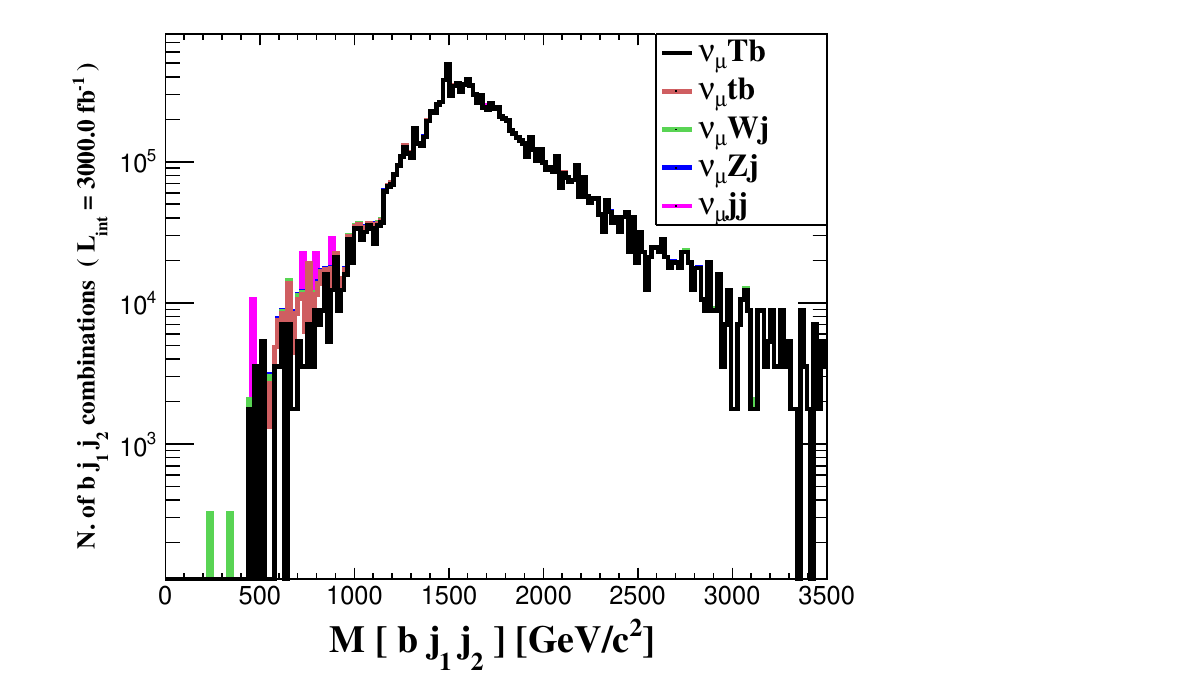}
\caption{}
\end{subfigure}\hspace{0.04\textwidth}
      \begin{subfigure}[t]{0.24\textwidth}
\centering
\includegraphics[width=\linewidth]{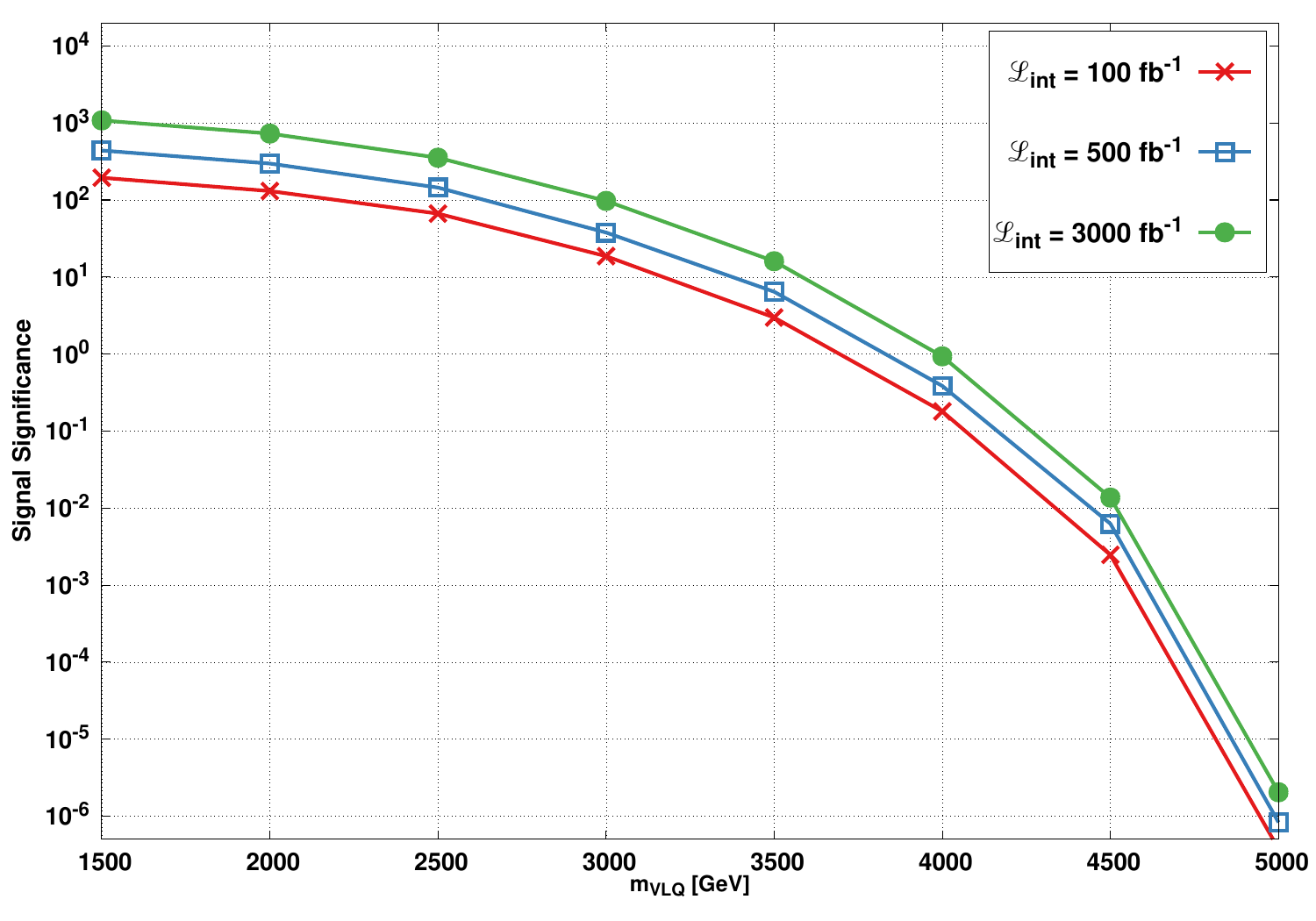}
\caption{}
\end{subfigure}
\caption{a) Reconstructed invariant mass $m_{bjj}$ showing the resonance peak for a 1.5~TeV VLQ-T. (b) Signal significance as a function of $m_{VLQ}$ for three integrated luminosity benchmarks at $\sqrt{s} = 9.16$~TeV.}
  \label{fig:invmass}
\end{figure*}

\section{Leptonic Analysis}
For the leptonic channel ($T\rightarrow Wb\rightarrow l\nu b$), Tables IV and V give the cumulative efficiencies. The efficiency is stable up to 4~TeV, since the high-$p_T$ lepton from the boosted $W$ is a robust trigger free of jet merging, and $e\nu$ and $\mu\nu$ perform comparably, allowing their combination. The $P_T(b)<700$~GeV cut removes extremely hard QCD $b$-jets.

\begin{table}[tbp]
\scriptsize\setlength{\tabcolsep}{2pt}\renewcommand{\arraystretch}{0.85}
\centering
\caption{Cumulative selection efficiencies for different $M_T$ values (GeV) corresponding to various kinematic selection cuts.}
\resizebox{\columnwidth}{!}{%
\begin{tabular}{|l|c|c|c|c|c|c|c|c|}
\hline
\textbf{Selection Cuts} & \textbf{$m_T=1500$} & \textbf{$m_T=2000$} & \textbf{$m_T=2500$} & \textbf{$m_T=3000$} & \textbf{$m_T=3500$} & \textbf{$m_T = 4000$} & \textbf{$m_T=4500$} & \textbf{$m_T=5000$} \\ 
&GeV&GeV&GeV&GeV&GeV&GeV&GeV&GeV \\
\hline
$\eta(l) > -6$ & 0.99332 & 0.99162 & 0.99276 & 0.99204 & 0.99107 & 0.99100 & 0.99195 & 0.98886 \\ 
$\eta(l) < 4$ & 0.99324 & 0.99162 & 0.99276 & 0.99196 & 0.99098 & 0.99087 & 0.99190 & 0.98886 \\ 
$P_T(l) > 100 GeV$ & 0.85273 & 0.90692 & 0.93126 & 0.94835 & 0.94703 & 0.95489 & 0.96203 & 0.96251 \\ 
$N(l) > 1.0$ & 0.85273 & 0.90692 & 0.93126 & 0.94835 & 0.94703 & 0.95489 & 0.96203 & 0.96251 \\ 
$\eta(b) > -2$ & 0.85186 & 0.90619 & 0.93086 & 0.94811 & 0.94700 & 0.95489 & 0.96203 & 0.96251 \\ 
$\eta(b) < 5$ & 0.84025 & 0.89410 & 0.91870 & 0.93814 & 0.93789 & 0.94950 & 0.96203 & 0.96251 \\ 
$P_T(b) < 700 GeV$ & 0.72901 & 0.53639 & 0.45459 & 0.40734 & 0.38534 & 0.36712 & 0.35443 & 0.34448 \\ 
$N(b) > 1.0$ & 0.64071 & 0.44271 & 0.35391 & 0.30552 & 0.27362 & 0.25114 & 0.23089 & 0.19453 \\ 
$N(b) < 3.0$ & 0.61920 & 0.42752 & 0.34274 & 0.29743 & 0.26694 & 0.24658 & 0.22734 & 0.19250 \\ 
MET $> 200.0$ & 0.36496 & 0.27386 & 0.24398 & 0.22380 & 0.21249 & 0.20137 & 0.19114 & 0.16413 \\ 
$\Delta R(l[1], b[1]) > 2.8$ & 0.30258 & 0.20666 & 0.16982 & 0.14710 & 0.13148 & 0.12237 & 0.11013 & 0.08328 \\ 
$\Delta R(b[1], b[1]) < 3.8$ & 0.26896 & 0.15172 & 0.10140 & 0.08071 & 0.07258 & 0.07306 & 0.06684 & 0.05309 \\  \hline
Total Events  & 99784 & 99953 & 99924 & 99828 & 100048 & 100040 & 100208 & 100039 \\ 
($\mathcal{L}_{int} = 3000 \text{ fb}^{-1}$) &   &  &  &  &  &  &  &  \\   \hline
\end{tabular}
}
\end{table}
\begin{table}[tbp]
\scriptsize\setlength{\tabcolsep}{3pt}\renewcommand{\arraystretch}{0.85}
\centering
\caption{Selection cuts applied to $\mu \nu$ and $e\nu$ channels.}
\begin{tabular}{|c|c|c|}
\hline
\textbf{Selection Cuts} & \textbf{$\mu \nu$} & \textbf{$e\nu$} \\ \hline
$\eta(l) > -6$ & 0.996572 & 0.994983 \\ 
$\eta(J) < 4$ & 0.990133 & 0.991406 \\ 
$P_{T}(l) > 100 GeV$ & 0.029709 & 0.34798 \\ 
$N(l) > 1.0$ & 0.029709 & 0.34798 \\ 
$\eta(b) > -2$ & 0.023826 & 0.31393 \\ 
$\eta(b) < 5$ & 0.023826 & 0.31393 \\ 
$P_{T}(b) > 700 GeV$ & 0.023806 & 0.30410 \\ 
$N(b) > 1.0$ & 0.021026 & 0.072820 \\ 
$N(b) < 3.0$ & 0.020739 & 0.072686 \\ 
MET $> 200.0$ & 2.01e-03 & 0.03018 \\ 
$\Delta R (l[1], b[1]) > 2.8$ & 3.07e-04 & 0.018252 \\ 
$\Delta R (b[1], b[1]) < 3.8$ & 3.07e-04 & 0.017495 \\ \hline
\textbf{Total Events} & 150004 & 149355 \\ \hline
\end{tabular}
\end{table}

Figure 9 shows a hard, central signal lepton (peak $>500$~GeV, a Jacobian peak smeared by the $W$ boost) with exactly one lepton selected. In Fig.~10 the signal neutrino gives a broad high-$E_T^{miss}$ peak, well separated from QCD. Figure 11 shows that requiring a $b$-tag rejects $W^{+}$+light-jet events and that signal $b$-jets are central and energetic. In Fig.~12 the signal (black line) has smaller lepton--$b$-jet $\Delta R$ than backgrounds because the high boost collimates the decay products \cite{Abdesselam}.
\begin{figure*}[t]
    \centering
  \begin{subfigure}[t]{0.27\textwidth}
\centering
\includegraphics[width=\linewidth]{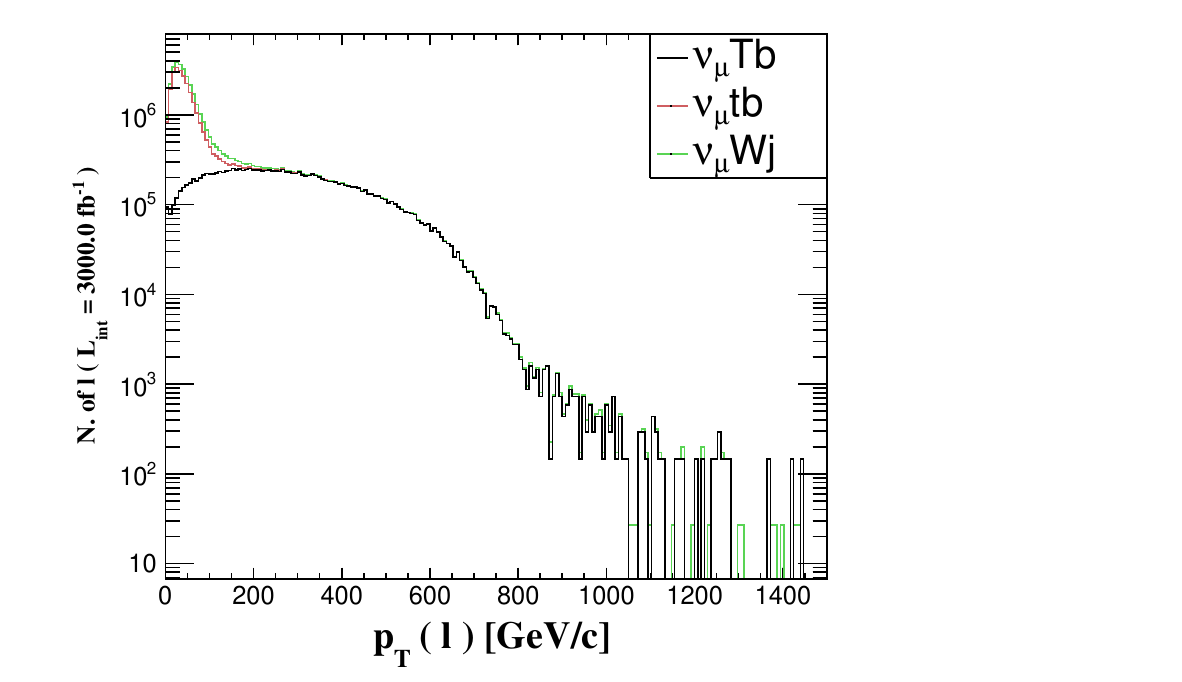}
\caption{}
\end{subfigure}\hfill
   \begin{subfigure}[t]{0.27\textwidth}
\centering
\includegraphics[width=\linewidth]{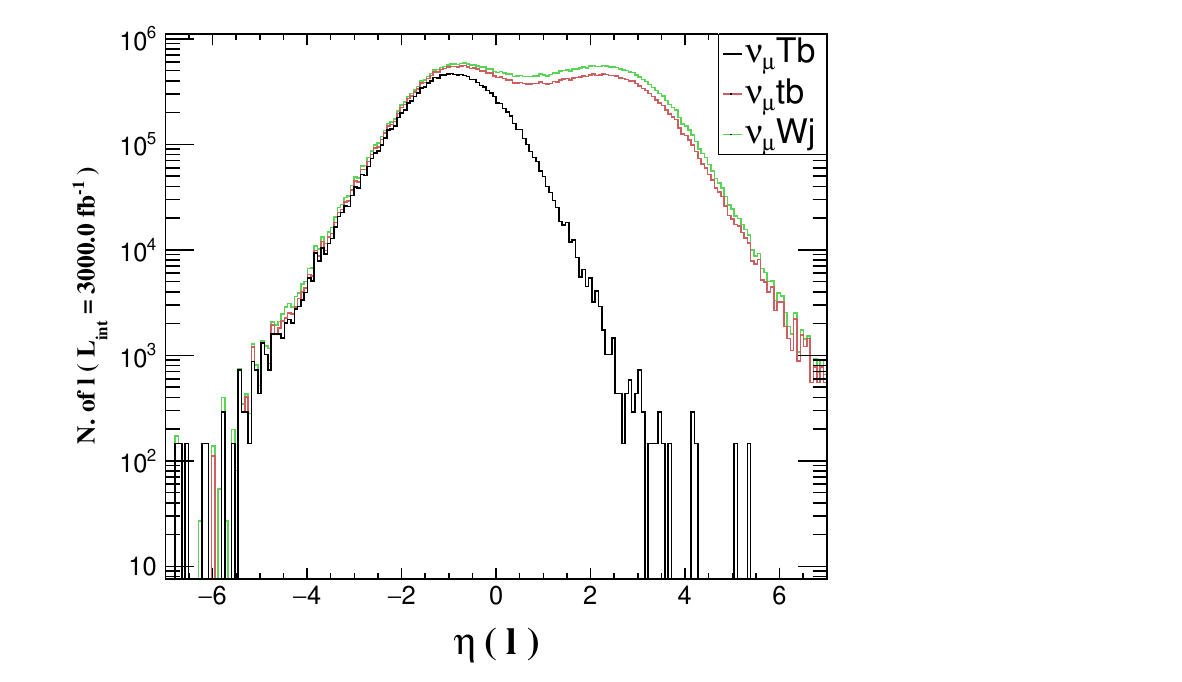}
\caption{}
\end{subfigure}\hfill
    \begin{subfigure}[t]{0.27\textwidth}
\centering
\includegraphics[width=\linewidth]{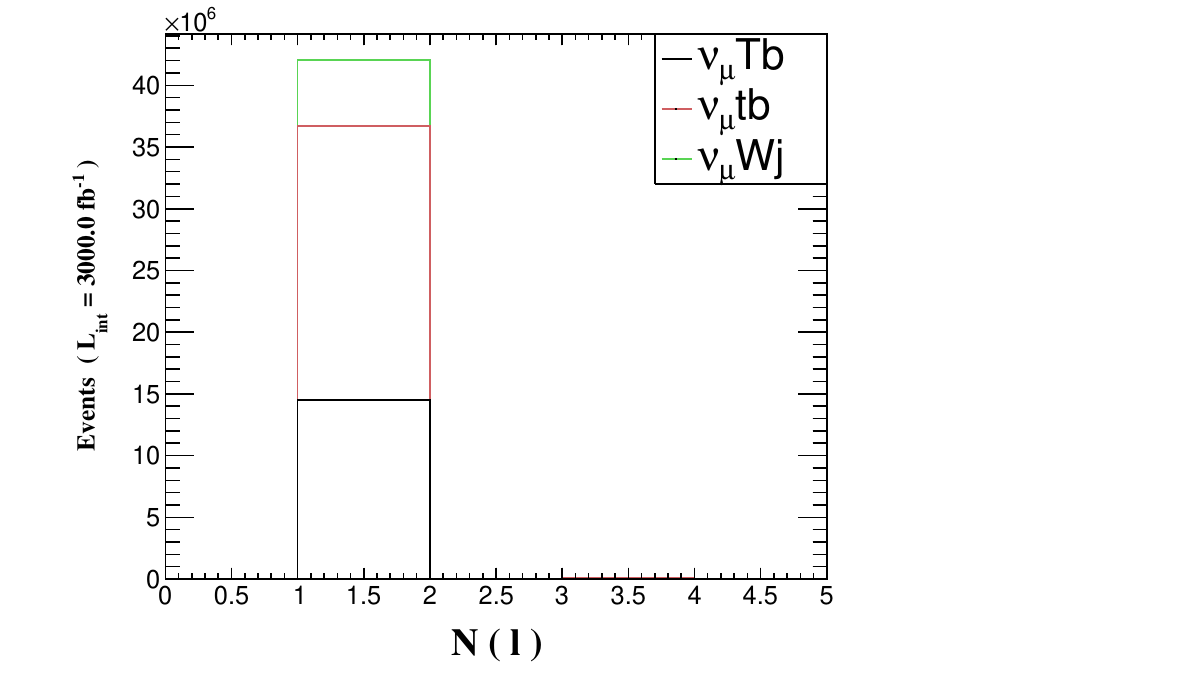}
\caption{}
\end{subfigure}
    \caption{Leptonic channel kinematics: (a) Lepton $p_T$ distribution showing the Jacobian peak from boosted $W$ decays. (b) Lepton pseudorapidity $\eta(l)$. (c) Lepton multiplicity ensures single-lepton selection criteria.}
    \label{fig:ptl-lep}
\end{figure*}
\begin{figure}[t]
    \centering
        \includegraphics[width=0.55\columnwidth]{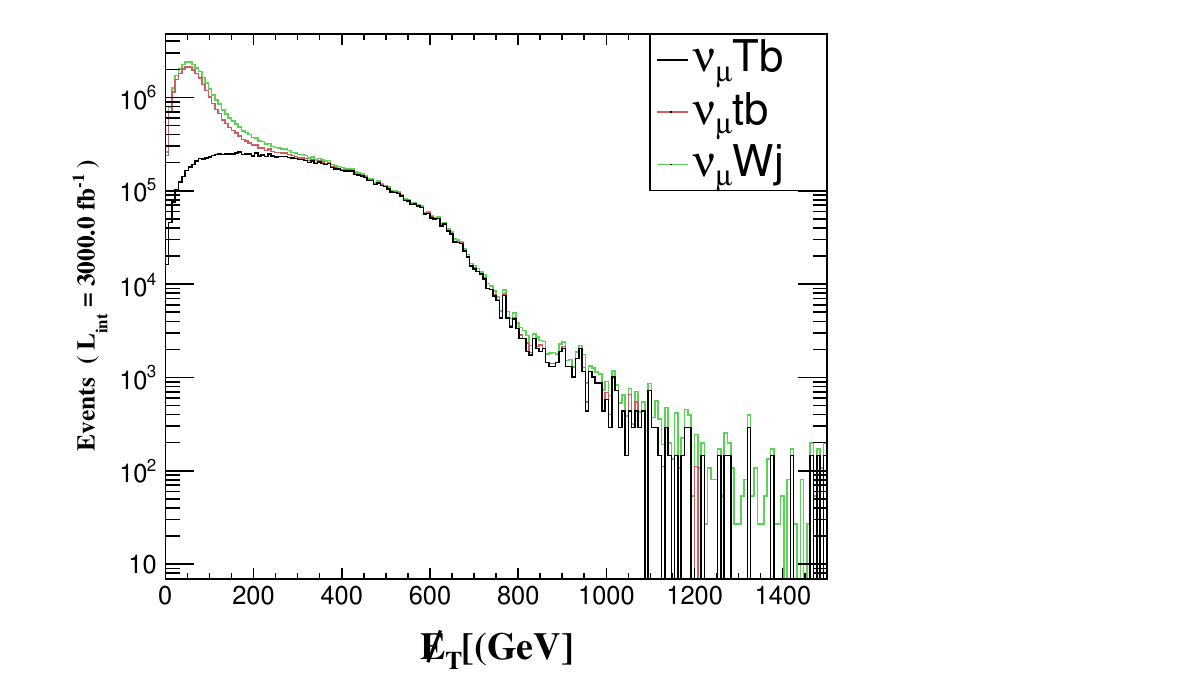}
    \caption{Missing transverse energy ($E_T^{miss}$) distribution in the leptonic channel, where the signal neutrino contributes to a high-energy tail compared to SM backgrounds.}
    \label{fig:et-lep}
\end{figure}
\begin{figure*}[t]
    \centering
  \begin{subfigure}[t]{0.27\textwidth}
\centering
\includegraphics[width=\linewidth]{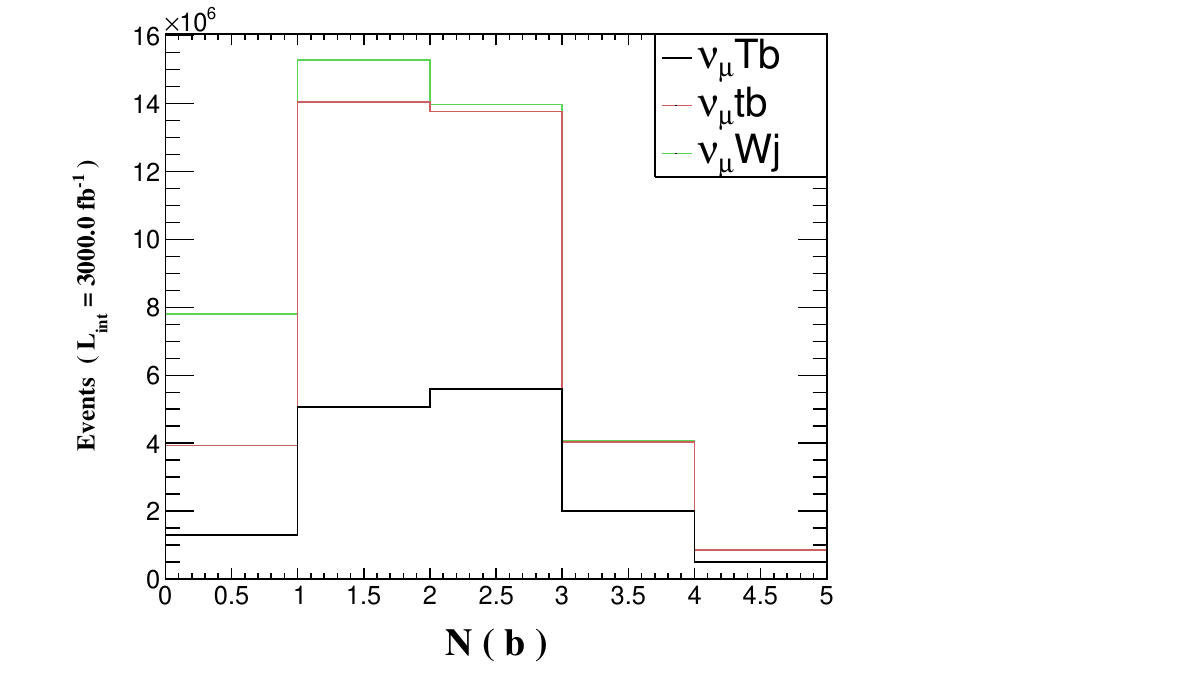}
\caption{}
\end{subfigure}\hfill
        \begin{subfigure}[t]{0.27\textwidth}
\centering
\includegraphics[width=\linewidth]{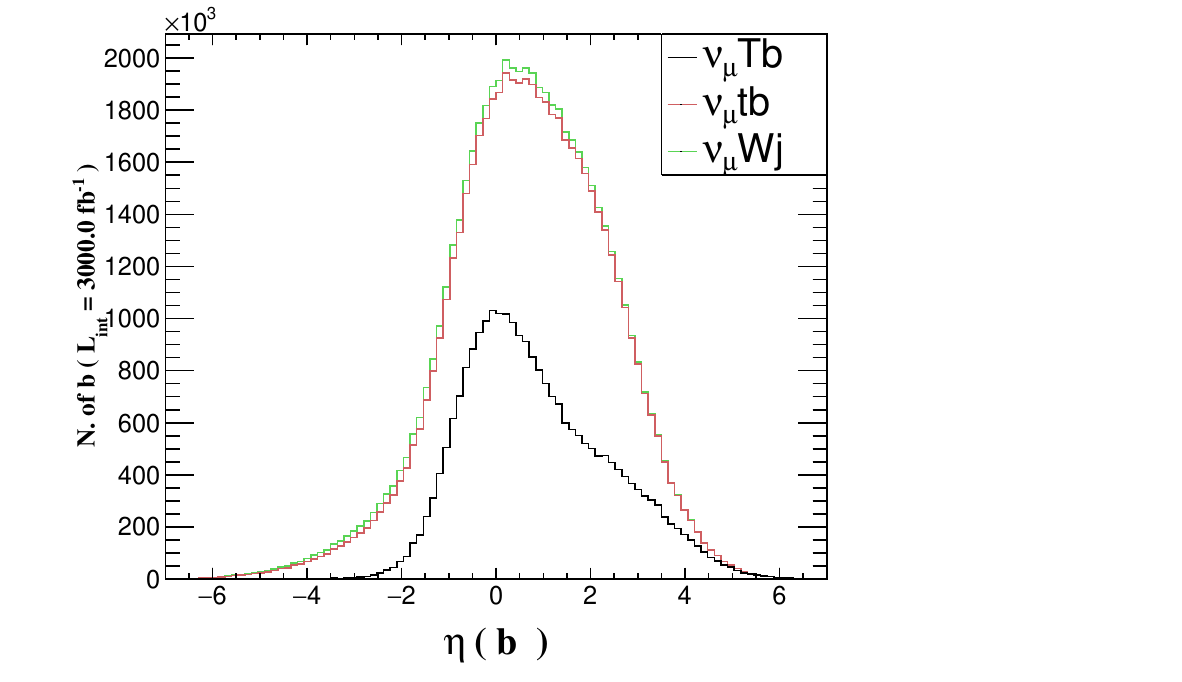}
\caption{}
\end{subfigure}\hfill
        \begin{subfigure}[t]{0.27\textwidth}
\centering
\includegraphics[width=\linewidth]{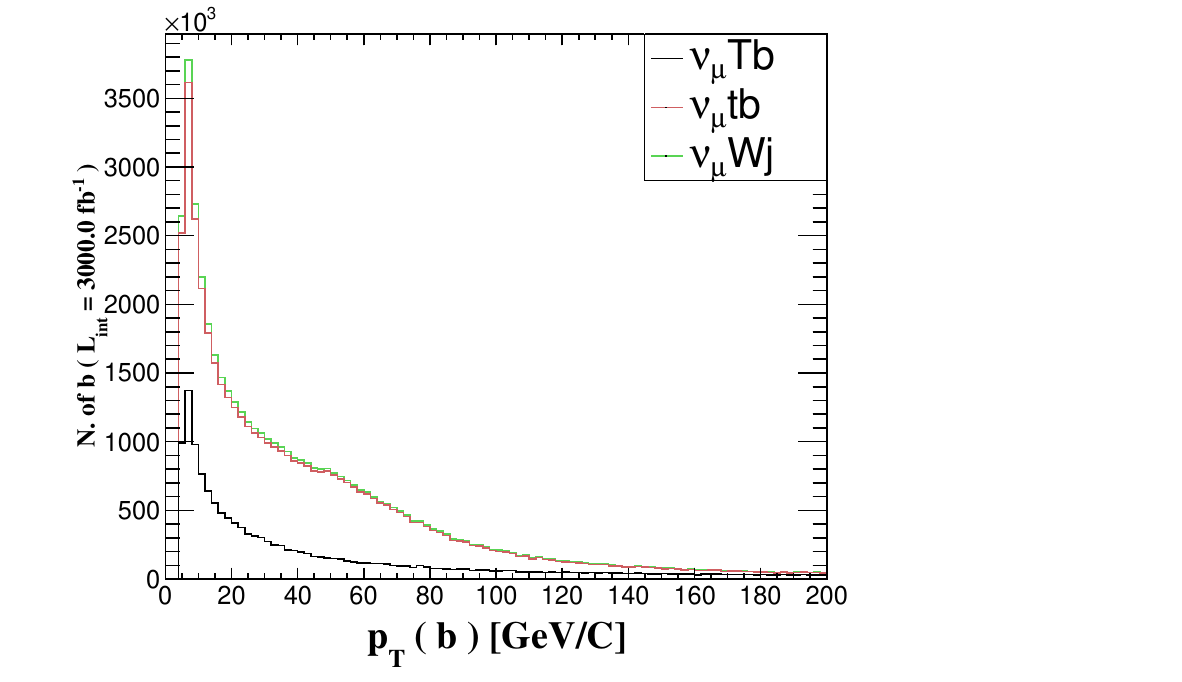}
\caption{}
\end{subfigure}
   \caption{B-tagging performance in the leptonic channel: (a) Number of b-tagged jets. (b) Pseudorapidity of b-jets. (c) Transverse momentum distribution of b-jets, showing the high-energy nature of signal b-quarks.}
   \label{fig:nb-lep}
\end{figure*}
\begin{figure*}[t]
    \centering
        \begin{subfigure}[t]{0.27\textwidth}
\centering
\includegraphics[width=\linewidth]{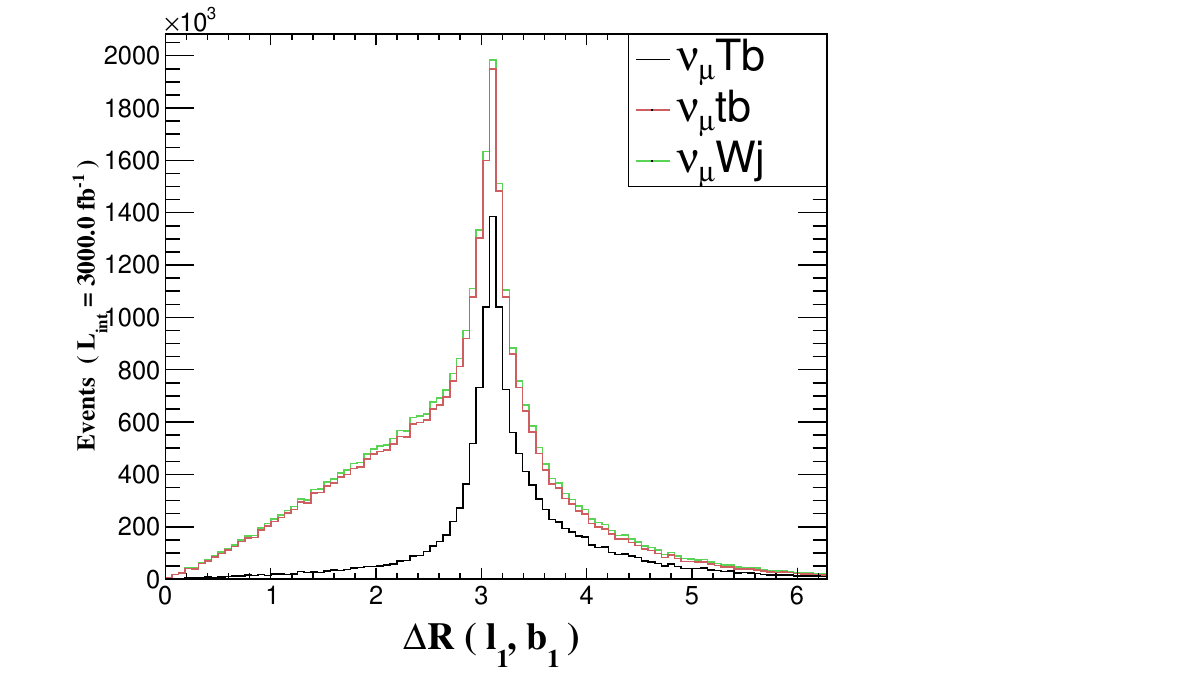}
\caption{}
\end{subfigure}\hspace{0.04\textwidth}
       \begin{subfigure}[t]{0.27\textwidth}
\centering
\includegraphics[width=\linewidth]{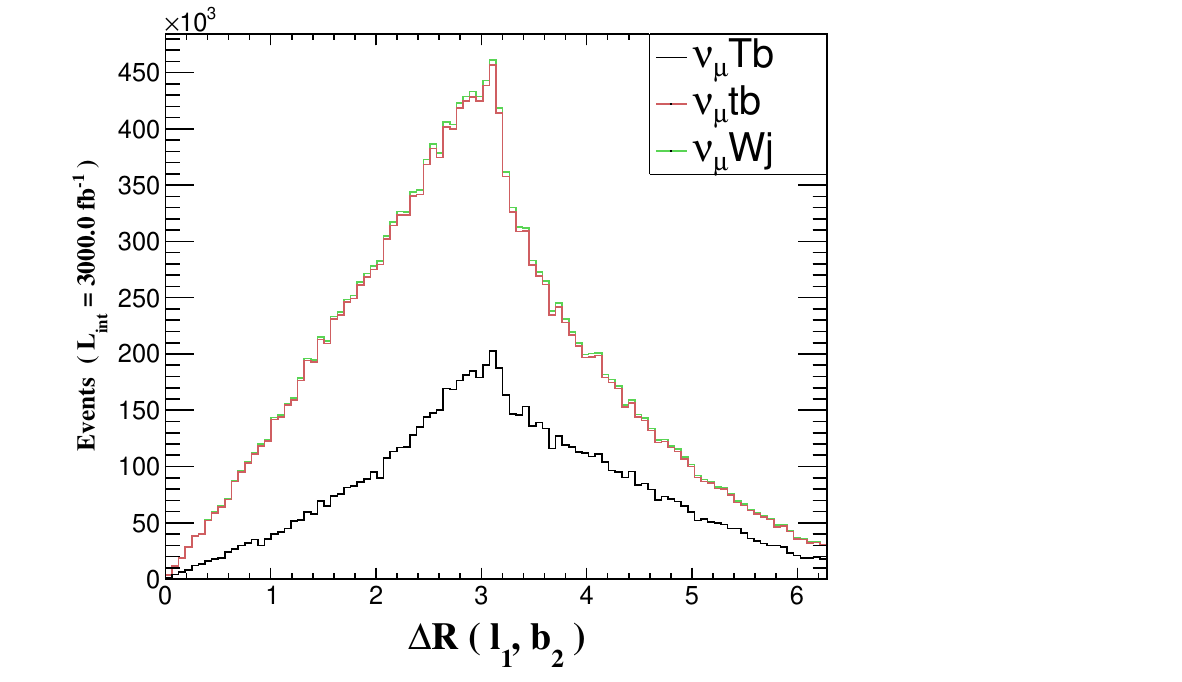}
\caption{}
\end{subfigure}
    \caption{Angular separation $\Delta R$ in the leptonic channel: (a) $\Delta R$ between the lepton and leading b-jet. (b) $\Delta R$ between the lepton and subleading b-jet. The smaller $\Delta R$ values at high mass indicate the collimated 'boosted' topology of the decay products.}
    \label{fig:deltar-lb1-lep}
\end{figure*}
Table VI lists the leptonic signal and background events: the rate is lower than in the hadronic channel but the background much smaller. Table VII compares the cross-sections: at $m_T=1.5$~TeV, $\sqrt{s}=5.29$~TeV and 500~fb$^{-1}$ the hadronic value (195.04 fb) exceeds the leptonic (119.828 fb), reflecting BR$(W\to q\bar q)\approx67\%$ versus BR$(W\to l\nu)\approx11\%$ per flavor, modified by channel-specific selections. The leptonic channel has higher selection efficiency (Table V) and lower background, so it is a critical cross-check of a hadronic discovery at $m_T\approx3$~TeV.

\begin{table}[tbp]
\scriptsize\setlength{\tabcolsep}{2pt}\renewcommand{\arraystretch}{0.85}
\caption{Signal and background event numbers with various significance metrics. \{Event numbers are scaled to a luminosity of $3000 \text{ fb}^{-1}$.}
\centering
\resizebox{\columnwidth}{!}{%
\begin{tabular}{|c|c|c|c|c|c|c|}
\hline
\textbf{$m_T$} & \textbf{Signal} & \textbf{Background} & \textbf{S/B} & \textbf{S/$\sqrt{B}$} & \textbf{S/(S+B)} & \textbf{S/$\sqrt{S+B}$} \\
\textbf{GeV} & \textbf{Events} & \textbf{Events} & & & &  \\
\hline
1500 & 16742 & 2724.6 & 6.145 & 318.37904 & 0.858 & 119.828 \\
2000 & 4671.0 & 2680.3 & 1.7427 & 88.94202 & 0.628 & 53.835 \\
2500 & 1122.7 & 2720.2 & 0.4127 & 20.22891 & 0.285 & 17.501 \\
3000 & 194.8 & 2879.4 & 0.06764 & 3.74804 & 0.0675 & 3.617  \\
3500 & 24.21 & 2718.3 & 0.00891 & 0.47503 & 0.00886 & 0.4643  \\
4000 & 1.58  & 2780.7 & 0.000570 & 0.029793 &  0.000547 & 0.0301  \\
4500 & 0.026 & 2732.5 & 9.51e-06 & 5.15e-04 & 1.01e-05  & 0.000498 \\
5000 & 5.25e-06 & 2824.8 & 1.86e-09 & 1.05e-07 & 1.93e-09 &  9.93e-08 \\
\hline
\end{tabular}
}
\label{tab:signal-background}
\end{table}
\begin{table}[tbp]
\scriptsize\setlength{\tabcolsep}{2pt}\renewcommand{\arraystretch}{0.85}
\caption{Cross sections (in fb) at $\sqrt{s} = 5.29, 6.48,$ and $9.16$ TeV, for Hadronic and Leptonic channels at different $m_T$ values at three different center-of-mass energies as mentioned above.}
\resizebox{\columnwidth}{!}{%
\begin{tabular}{|c|c|c|c|c|c|c|c|c|c|}
\hline
\textbf{Energy} & \textbf{Channel} & \textbf{$m_T$=1500} & \textbf{$m_T$=2000} & \textbf{$m_T$=2500} & \textbf{$m_T$=3000} & \textbf{$m_T$=3500} & \textbf{$m_T$=4000} & \textbf{$m_T$=4500} & \textbf{$m_T$=5000} \\ 
 $\sqrt{s}$  & & GeV&GeV & GeV&GeV & GeV& GeV& GeV&GeV\\ 
\hline
5.29 TeV & Hadronic & 195.04 & 130.700 & 66.660 & 18.666  & 2.985 & 0.180 &0.00247&3.76e-07 \\
         & Leptonic & 119.82 & 53.835 & 17.501 & 3.617 & 0.4643 & 0.0301 & 0.000498 & 9.93e-08 \\   \hline
6.48 TeV & Hadronic & 320.53  & 255.18  & 177.18  & 98.228	 & 39.378	 & 11.349 & 1.772  & 0.1357 \\
        & Leptonic & 186.14 & 106.919 & 53.470  & 22.7 & 7.407  & 1.700 & 0.2722   & 0.024 \\
\hline 
9.16 TeV & Hadronic & 594.93 & 575.48 & 502.84 & 424.96 & 325.52 & 230.47 & 136.9 & 70.144\\ 
         & Leptonic & 334.94 & 239.11 & 162.56  & 110.23   & 76.736   & 43.387  & 24.50 & 11.356 \\ \hline
\end{tabular}
}
\end{table}

\section{Conclusion}
We studied singly produced $T\to Wb$ quarks at a future muon-proton collider; high-$p_T$ central jets, isolated leptons and large $H_T$ discriminate the signal from SM backgrounds. At $\sqrt{s}=9.16$~TeV and 3000~fb$^{-1}$ the hadronic channel ($T \to Wb \to jjb$) gives about 21.8$\sigma$ for $m_T=3$~TeV (Table 3), while the leptonic channel ($T \to Wb \to l\nu b$) has a lower rate but much lower background, with about 3.7$\sigma$ (Table 6); the 1.5~TeV hadronic cross-section is 594.93~fb (Table 7). The significance curves (Fig.~8(b)) and machine-learning comparison (Fig.~13) maintain the $5\sigma$ threshold up to about 3.5~TeV, beyond which the cross-section, which scales quadratically with the coupling, becomes negligible. A 9.16~TeV $\mu p$ collider with 3000~fb$^{-1}$ can thus discover singly produced $T$ quarks up to approximately 3.5~TeV, well beyond current limits and the HL-LHC, combining hadronic statistics with a clean leptonic cross-check (Tables 5, 7). BDT/neural-network analyses and profile likelihood fits of $H_T$ could extend the reach further.

\begin{table}[tbp]
\scriptsize\setlength{\tabcolsep}{2pt}\renewcommand{\arraystretch}{0.85}
\centering
\caption{Distribution of the cumulative signal efficiencies for the hadronic channel across all $m_{T}$ (2000-5000 GeV) points at $\sqrt{s} = 5.29$ TeV.}
\label{tab:tableVIII}
\resizebox{\columnwidth}{!}{%
\begin{tabular}{|c|c|c|c|c|c|c|c|}
\hline
\textbf{Selection Cuts}& \textbf{$m_T$=2000} & \textbf{$m_T$=2500} & \textbf{$m_T$=3000} & \textbf{$m_T$=3500} & \textbf{$m_T$=4000} & \textbf{$m_T$=4500} & \textbf{$m_T$=5000} \\
\textbf{(GeV)} & \textbf{(GeV)} & \textbf{(GeV)} & \textbf{(GeV)} & \textbf{(GeV)} & \textbf{(GeV)} &  \textbf{(GeV)}& \textbf{(GeV)} \\
\hline
$|\eta_j| < 2.5$ & 0.50068 & 0.49836 & 0.49784 & 0.50288 & 0.49940 & 0.50016 & 0.49928 \\
\hline
$p_T^{j_1} > 120\,\mathrm{GeV}$ & 0.49532 & 0.49472 & 0.49508 & 0.50088 & 0.49808 & 0.49872 & 0.49828 \\
\hline
$p_T^{j_2} > 50\,\mathrm{GeV}$ & 0.49084 & 0.48988 & 0.48824 & 0.49092 & 0.48456 & 0.48232 & 0.47168 \\
\hline
$N_{\mathrm{jets}} \ge 4$ & 0.17828 & 0.16952 & 0.16064 & 0.15420 & 0.14428 & 0.13532 & 0.09748 \\
\hline
$N_{b\mathrm{jets}} \ge 2$ & 0.05388 & 0.04720 & 0.03932 & 0.03288 & 0.02980 & 0.02548 & 0.01600 \\
\hline
$H_T > 450\,\mathrm{GeV}$ & 0.05376 & 0.04712 & 0.03924 & 0.03288 & 0.02968 & 0.02548 & 0.01600 \\
\hline
$E_T^{\mathrm{miss}} > 10\,\mathrm{GeV}$ & 0.05316 & 0.04640 & 0.03888 & 0.03276 & 0.02948 & 0.02532 & 0.01596 \\
\hline
\textbf{$\epsilon$ (\%)} & \textbf{5.316} & \textbf{4.640} & \textbf{3.888} & \textbf{3.276} & \textbf{2.948} & \textbf{2.532} & \textbf{1.596} \\
\hline
\end{tabular}
}%
\end{table}
\begin{table}[htbp]
\scriptsize\setlength{\tabcolsep}{2pt}\renewcommand{\arraystretch}{0.85}
  \centering
  \caption{Hadronic optimized performance at $\sqrt{s}=9.16~\mathrm{TeV}$ and $m_T=3000~\mathrm{GeV}$.}
  \label{tab:IX}
  \resizebox{\columnwidth}{!}{%
  \begin{tabular}{|c|c|c|c|c|c|c|}
    \hline
    Luminosity ($\mathrm{fb}^{-1}$) & BDT cut & BDT $S/B$ & BDT $S/\sqrt{S+B}$ & MLP cut & MLP $S/B$ & MLP $S/\sqrt{S+B}$ \\
    \hline
    100  & 0.196 & 15.14 & 49.39  & 0.900 & 39.63 & 49.94  \\ \hline
    500  & 0.196 & 15.14 & 110.45 & 0.900 & 39.63 & 111.66 \\ \hline
    3000 & 0.196 & 15.14 & 270.54 & 0.900 & 39.63 & 273.51 \\  \hline
    \hline
  \end{tabular}
  }
\end{table}

\section{Machine-learning Based Discrimination Method} 
\begin{figure}[t]
    \centering
    \includegraphics[width=0.66\columnwidth]{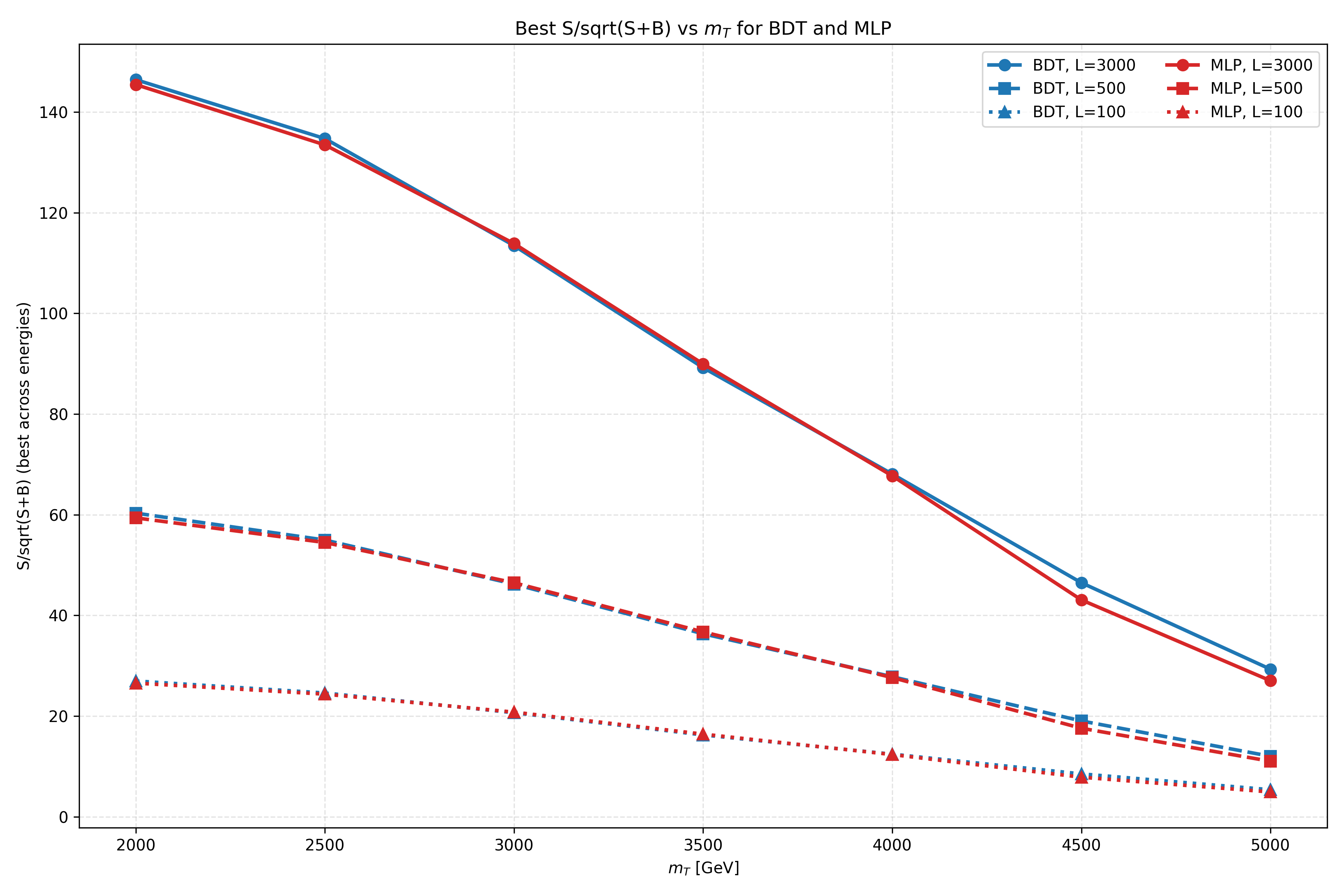}
    \caption{A view of the $m_{T}$ scan as a function of signal significance at different integrated luminosities for two different classifiers.}
    \label{fig:fig13}
\end{figure}
We optimize the selection with Boosted Decision Trees (BDT) and Multi-Layer Perceptrons (MLP) \cite{Roe:2004na, Hocker:2007ht,Baldi:2014kfa} in both channels at $\sqrt{s}=9.16$~TeV and $m_T=3000$~GeV. Hadronic inputs are $H_T$, jet $p_T^j,\eta^j$, $N_j,N_b$ and $\Delta R(b_1,b_2)$; leptonic inputs are $p_T^l,\eta^l$, $E_T^{miss}$, $p_T^b,\eta^b$ and $\Delta R(l,b)$. The MLP has three hidden layers (ReLU), a sigmoid output, the Adam optimizer (learning rate 0.001) and binary cross-entropy loss. Thresholds were optimized for $S/B$ and $S/\sqrt{S+B}$ \cite{Cowan}; Fig.~13 compares the classifiers.

The MLP outperforms the BDT in both channels at all luminosities; in the hadronic channel (Table~\ref{bdtmlp:had}) its $S/B$ is larger by a factor of about 2.62 with a small gain in $S/\sqrt{S+B}$, nearly independent of luminosity, reflecting its ability to capture correlations in high-multiplicity environments \cite{Baldi:2014kfa,Radovic:2018dip}.

\begin{table}[htbp]
\scriptsize\setlength{\tabcolsep}{2pt}\renewcommand{\arraystretch}{0.85}
  \centering
  \caption{Relative MLP improvement over BDT in the hadronic channel.}
  \label{tab:tableX}
  \resizebox{\columnwidth}{!}{%
\begin{tabular}{|c|c|c|}
 \hline
    Luminosity ($\mathrm{fb}^{-1}$) & $S/B$ improvement & $S/\sqrt{S+B}$ improvement \\
    \hline
    100  & +161.76\% & +1.11\% \\  \hline
    500  & +161.76\% & +1.10\% \\   \hline
    3000 & +161.76\% & +1.10\% \\   \hline
    \hline
  \end{tabular}
}
  \label{bdtmlp:had}
\end{table}
\begin{table}[htbp]
\scriptsize\setlength{\tabcolsep}{2pt}\renewcommand{\arraystretch}{0.85}
  \centering
  \caption{Leptonic optimized performance at $\sqrt{s}=9.16~\mathrm{TeV}$ and $m_T=3000~\mathrm{GeV}$.}
  \label{tab:tableXI}
  \resizebox{\columnwidth}{!}{%
  \begin{tabular}{|c|c|c|c|c|c|c|}
    \hline
    Luminosity ($\mathrm{fb}^{-1}$) & BDT cut & BDT ($S/B$) & BDT ($S/\sqrt{S+B}$) & MLP cut & MLP ($S/B$) & MLP ($S/\sqrt{S+B}$) \\
    \hline
    100  & 0.240 & 10.89 & 20.67  & 0.925 & 11.93 & 20.79  \\ \hline
    500  & 0.240 & 10.89 & 46.21  & 0.925 & 11.93 & 46.48  \\ \hline
    3000 & 0.214 & 10.93 & 113.45 & 0.925 & 11.93 & 113.85 \\ \hline
      \end{tabular}
  }
\end{table}
\begin{table}[tbp]
\scriptsize\setlength{\tabcolsep}{2pt}\renewcommand{\arraystretch}{0.85}
  \centering
  \caption{Relative MLP improvement over BDT in the leptonic channel compared with hadronic, the leptonic channel exhibits a more modest but consistent MLP advantage. The purity improvement remains near the 10\% level, and the significance gain is positive for all luminosity points, indicating no degradation from the higher-threshold MLP working point.}
  \label{tab:tableXII}
  \resizebox{\columnwidth}{!}{%
\begin{tabular}{|c|c|c|}
   \hline
    Luminosity ($\mathrm{fb}^{-1}$) & Improvement ($S/B$)  & Improvement ($S/\sqrt{S+B}$)  \\
 \hline
    100  & +9.55\% & +0.58\% \\ \hline
    500  & +9.55\% & +0.58\% \\ \hline
    3000 & +9.15\% & +0.35\% \\  \hline
      \end{tabular}
}
\end{table}

Table~\ref{tab:tableVIII} gives the hadronic signal efficiencies at 5.29~TeV. Tables~\ref{tab:IX} and \ref{tab:tableX} show the higher MLP $S/B$ in the hadronic sector, and Tables~\ref{tab:tableXI} and \ref{tab:tableXII} the same modest, consistent trend in the leptonic channel.

\section*{Declaration of Generative AI and AI-assisted technologies}
During the preparation of this work, the authors used generative AI tools to assist in verifying the bibliography and refining the style. The authors take full responsibility for the manuscript; the physics analysis, simulation and code were conducted entirely by the authors without AI assistance.

{\fontsize{6.5}{7.5}\selectfont

}

\end{document}